\documentclass[preprintnumbers,amsmath,amssymb,showpacs]{revtex4}
\usepackage{amsmath,amsfonts,amssymb,latexsym,graphicx,youngtab}
\newcommand{\dis}{\displaystyle}

\newcommand{\bequ}{\begin{equation}}
\newcommand{\eequ}{\end{equation}}
\newcommand{\barr}{\begin{array}}
\newcommand{\earr}{\end{array}}
\newcommand{\bea}{\begin {eqnarray}}
\newcommand{\eea}{\end {eqnarray}}
\newcommand{\lb}{\label}

\renewcommand{\Im}{{\cal I}{\rm m}\:}

\begin{document}
\let\la=\lambda
\def \Z {\mathbb{Z}}
\def \Zt {\mathbb{Z}_o^4}
\def \R {\mathbb{R}}
\def \C {\mathbb{C}}
\def \La {\Lambda}
\def \ka {\kappa}
\def \vphi {\varphi}
\def \Zd {\Z ^d}
\title{Dynamical Eightfold Way in Strongly Coupled Lattice QCD}
\author{Paulo A. Faria da Veiga} \email
{veiga@icmc.usp.br}
\author{Michael O'Carroll}
\affiliation{Departamento de Matem\'atica Aplicada e
Estat\'{\i}stica, ICMC-USP,
\\
C.P. 668, 13560-970 S\~ao Carlos SP, Brazil}

\pacs{11.15.Ha, 02.30.Tb, 11.10.St, 24.85.+p\\\ \ Keywords: Lattice
QCD, Excitation Spectrum, Eightfold Way, Spectral Analysis}
\date{July 11, 2007\vspace{.3cm}}

\begin{abstract}
We obtain from first principles, i.e. from the quark-gluon dynamics,
the Gell'Mann-Ne'eman eightfold way baryons in an imaginary-time
functional integral formulation of $3+1$ lattice QCD in the strong
coupling regime (small hopping parameter $\kappa>0$). The model has
 ${\rm SU}(3)_c$ gauge and global ${\rm SU}(3)_f$ flavor
symmetries. In the subspace of the quantum mechanical physical
Hilbert space of vectors with an odd number of quarks, the baryons
are associated with isolated dispersion curves in the
energy-momentum spectrum. The spin $1/2$ octet and spin $3/2$
decuplet baryons have asymptotic mass $-3\ln\kappa$ and for each
baryon there is an antibaryon with identical spectral properties.
All the masses have the form
$M=-3\ln\kappa-3\kappa^3/4+\kappa^6r(\kappa)$, with $r(\kappa)$ real
analytic. For each member of the octet $r(\kappa)$ is the same; for
each member of the decuplet, $r(0)$ is the same. So, there is no
mass splitting within the octet, and within the decuplet up to and
including ${\cal O}(\kappa^6)$. However, there is an octet-decuplet
mass difference of $3\kappa^6/4+{\cal O}(\kappa^7)$. The baryon and
anti-baryon spectrum is the only spectrum up to near the
meson-baryon threshold of $\approx -5\ln\kappa$. A decoupling of
hyperplane method is used to naturally unveil the form of the baryon
composite fields (there is no a priori guesswork), to show the
existence of particles and their multiplicities using a spectral
representation for the two-baryon correlation. We also obtain the
(anti-)baryon dispersion curves which admit the representation
$w(\kappa,\vec
p)=-3\ln\kappa-3\kappa^3/4+\kappa^3\sum_{j=1,2,3}\,(1-\cos
p^j)/4+r(\kappa,\vec p)$, where $r(\kappa,\vec p)$ is of ${\cal
O}(\kappa^6)$. For the octet, $r(\kappa,\vec p)$ is jointly analytic
in $\kappa$ and in each $p^j$, for small $|\Im p^j|$.
\end{abstract}
\maketitle
In Ref. \cite{GellNee} a quark model with three flavors ({\em up}
${\rm u}$, {\em down} ${\rm d}$ and {\em strange} ${\rm s}$) and an
${\rm SU}(3)_f$ flavor symmetry was introduced to describe hadrons
by an eightfold way classification scheme. A dynamical ${\rm
SU}(3)_c$ local gauge model of quarks and gluons and color dynamics
was proposed later, the well-known QCD, as a model for the strong
interactions. It was shown to be asymptotically free \cite{GWP}, and
perturbation theory was used successfully for high energy phenomena
but not at low energies. To understand the low-lying energy-momentum
(E-M) spectrum and confinement (no isolated quarks are observed) a
lattice approximation in an imaginary-time functional integral
formulation was introduced in \cite{Wil2}. The use of this
approximation in different contexts, for example the strong coupling
expansion, can be found e.g. in Refs. \cite{Wil,Group,Creu2,MM}.
Numerical simulations on the lattice acquired an important status to
determine the particle content of the model and to give an answer to
other questions which were not attainable using perturbation theory
(see \cite{num}).

In a mathematically rigorous treatment, and in an imaginary-time
setting, a physical Hilbert space ${\cal H}$ and E-M operators are
constructed for the lattice QCD in \cite{OS,Sei}. A Feynman-Kac
(F-K) formula is also established.

In a series of papers in Refs. \cite{CMP,ours,longo}, we determined
the low-lying E-M spectrum of increasingly complex ${\rm SU}(3)_c$
lattice QCD models in the strong coupling regime, i.e. with small
hopping parameter $\kappa>0$ and plaquette coupling
$\beta=1/(2g_0^2)$ (large glueball mass) obeying $\beta\ll\kappa\ll
1$; we obtained the one-hadron and the two-hadron bound-state
spectra, up to the two-particle energy threshold.

Here, we obtain the baryon part of the spectrum in the more
realistic ${\rm SU}(3)_c$ lattice QCD model with three flavors, in
$3+1$ dimensions and in the strong coupling regime. We derive the
Gell'Mann-Ne'eman eightfold way baryons exclusively from the
quark-gluon dynamics. No guesswork is needed regarding the form of
the baryon composite fields. We show the existence of $56$ baryons
and their anti-particles. The baryons have asymptotic mass $\approx
-3\ln\kappa$ and form the spin $1/2$ octet and the spin $3/2$
decuplet of the eightfold way. Anti-baryons and baryons have the
same spectral properties by charge conjugation, and we show that
they give rise to {\em all} the E-M spectrum in the subspace ${\cal
H}_o\subset{\cal H}$ of vectors with an {\em odd} number of quarks,
up to near the meson-baryon threshold of $\approx -5\ln\kappa$. The
other eightfold way particles are mesons, which lie in the even
sector of ${\cal H}$, and have asymptotic masses $\approx
-2\ln\kappa$. The eightfold way mesons are treated in Ref.
\cite{8fdmesons}.

Our lattice QCD model has the partition function
$Z=\int\,e^{-S(\psi,\bar \psi,g)}\,d\psi\,d\bar\psi\,d\mu(g)\,,$ and
for $F(\bar\psi, \psi, g)$, the normalized correlations are denoted
by $\langle \,F\,\rangle=\frac1{Z}\int\,F(\bar\psi, \psi,
g)\,e^{-S(\psi,\bar
 \psi,g)}\,d\psi\,d\bar\psi\,d\mu(g)$.
The gauge-invariant action $S\equiv S(\psi,\bar\psi,g)$ is Wilson's
action \cite{Wil}
 with an ${\rm SU}(3)_f$ flavor symmetry given by\bequ \barr{lll}
S&=&\frac{\kappa}{2}\sum\,\bar\psi_{a,\alpha,f}(u)\Gamma^{\sigma
e^\mu}_{\alpha\beta}(g_{u,u+\sigma e^\mu})_{ab}
\psi_{b,\beta,f}(u+\sigma e^\mu)
+\sum_{u\in\Zt} \,\bar\psi_{a,\alpha,f}(u)M_{\alpha\beta}
\psi_{a,\beta,f}(u)-\frac {1}{g_0^2}\sum_{p}\,\chi(g_p)\,,\earr
\lb{action}\eequ where, besides the sum over repeated indices
$\alpha,\:\beta=1,2,3,4$ (spin), $a=1,2,3$ (color) and
$f=1,2,3\equiv u,d,s$ (isospin), the first sum runs over
$u=(u^0,\vec u)=(u^0,u^1,u^2,u^3)\in \Zt\equiv\left\{\pm 1/2,\pm
3/2,\pm 5/2...\right\}\times \mathbb{Z}^3$, $\sigma=\pm1$ and
$\mu=0,1,2,3$. Here, $0$ denotes the time direction and the $3$
direction is also called the $z$-direction. $e^\mu$ is the unit
lattice vector for the $\mu$-direction. At a site $u\in\Zt$,
$\hat\psi_{a\alpha f}(u)$ are fermionic Grassmann fields (the upper
hat meaning the presence or absence of a bar) and we refer to
$\alpha=1,2$ as {\em upper} spin indices and $\alpha=3,4$
(equivalently, $+$ or $-$ respectively) as {\em lower} ones. For
each nearest neighbor oriented bond $< u,u\pm e^\mu>$ there is an
${\rm SU}(3)_c$ matrix $U(g_{u,u\pm e^\mu})$ parametrized by the
gauge group element $g_{u,u\pm e^\mu}$ and satisfying
$U(g_{u,u+e^\mu})^{-1}=U(g_{u+e^\mu,u})$. For simplicity, we
sometimes drop the $U$ from the notation. To each oriented plaquette
$p$ there is a plaquette variable $\chi(U(g_p))$ where $U(g_p)$ is
the orientation-ordered product of matrices of ${\rm SU}(3)_c$, and
$\chi$ is the real part of the trace. $M\equiv M(m,\kappa)
=m+2\kappa$ and, given $\kappa$, $m>0$ is chosen such that
$M_{\alpha\beta}= \delta_{\alpha\beta}$, so that
$m=1-2\kappa\lesssim 1$ in the strong coupling regime. Also, we take
$\Gamma^{\pm e^\mu}=-I_4\pm\gamma^\mu$, where $\gamma^0=
\left(\barr{cc} I_2&0
\\0&-I_2\earr\right)$, $\gamma^j= \left(\barr{cc} 0&i{\sigma}^j
\\-i{\sigma}^j&0\earr\right)$, $j=1,2,3$, are the
$4\times 4$ Dirac matrices and satisfy
$\{\gamma_\mu,\gamma_\nu\}=2\delta_{\mu\nu}I_4$, where
$\sigma^{j=1,2,3}$ are Pauli matrices. $d\mu(g)$ is the product
measure over non-oriented bonds of normalized ${\rm SU}(3)_c$ Haar
measures (see Ref. \cite{Si2}), $g_{uv}$ and $g^{-1}_{vu}$ are
indistinct integration variables. The Grassmann integrals are given
in \cite{Ber}; for $\kappa=0$, $\langle
\psi_{\ell_1}(x)\,\bar\psi_{\ell_2}(y)\rangle=\delta_{\alpha_1,\alpha_2}
\delta_{a_1a_2}\delta_{f_1f_2}\delta(x-y)$, and the integral of
monomials is given by Wick압 theorem. For more details about the
model definition and notation, see Refs. \cite{CMP,longo}. For free
fermions, there is no spectral doubling and the dispersion curve is
monotone increasing in momentum $\vec p$, convex for small $|\vec
p|$.

The physical quantum mechanical Hilbert space ${\cal H}$ and the E-M
operators $H$ and $P^j$, $j=1,2,3$, are defined as in \cite{OS,Sei,
CMP}. Polymer expansion methods \cite{Sei,Si} ensure the
thermodynamic limit of correlations exists and truncated
correlations have exponential tree decay. The limiting correlations
are lattice translational invariant and extend to analytic functions
in the global coupling parameters $\kappa$ and $\beta=1/(2g_0^2)$
and also in any finite number of local coupling parameters. For
gauge-invariant $F$ and $G$ restricted to $u^0=1/2$, we have the F-K
formula \bequ\lb{FeyKa}(G,\check T_0^{x^0}\check T_1^{x^1}\check
T_2^{x^2}\check T_3^{x^3}F)_{{\cal H}} =\langle [T_0^{x^0}\vec
T^{\vec x}F]\Theta G\rangle\,\,,\eequ where $T_0^{x^0}$,
$T_i^{x^i}$, $i=1,2,3$, denote translation of the functions of
Grassmann and gauge variables by $x^0\geq0$, $\vec
x=(x^1,x^2,x^3)\in\Z^{3}$, $T^{\vec x}=T_1^{x^1}T_2^{x^2} T_3^{x^3}$
and $\Theta$ is an antilinear, order reversing operator which
involves time reflection \cite{Sei}. For simplicity, in Eq.
(\ref{FeyKa}), we do not distinguish between Grassmann, gauge
variables (rhs) and their associated Hilbert space vectors (lhs) in
our notation. As linear operators in ${\cal H}$, $\check
T_{\mu=0,1,2,3}$ are mutually commuting; $\check T_0$ is
self-adjoint, with $-1\leq \check T_0\leq1$, and $\check
T_{j=1,2,3}$ are unitary. So, $\check T_j=e^{iP^j}$ defines the
self-adjoint momentum operator $\vec P=(P^1,P^2,P^3)$ with spectral
points $\vec p\in{\bf T}^3\equiv (-\pi,\pi]^3$ and $\check
T_0^2=e^{-2H}\geq0$ defines the energy operator $H\geq0$. We call a
point in the E-M spectrum with $\vec p=\vec 0$ a mass. Also, we let
${\cal E}(\la^0,\vec \la)$ be the product of the spectral families
of $\check T_0$, $P^1$, $P^2$ and $P^3$.

In order to classify and label the baryon states, we note that the
action of Eq. (\ref{action}) has a ${\rm SU}(2)\oplus {\rm SU}(2)$
spin symmetry at $\kappa=0$ in the hopping term, separately in the
lower and upper components. Motivated by this property, we introduce
spin operators $J_x$, $J_y$, $J_z$ and $\vec J^{\,2}\equiv
J_x^2+J_y^2+ J_z^2$ acting on the Grassmann field algebra, and which
obey the usual angular momentum commutation relations. Although we
adopt the terminology of the continuum, the analogy with the
continuum only holds for $\kappa=0$ for which the spin symmetry is
exact.

To show the existence of particles up to near the meson-baryon
threshold, we obtain a matrix valued two-point function $G(u,v)$
which has a spectral representation, derived by using the F-K
formula and the spectral representations of the E-M operators. The
Fourier transform $\tilde G(p)=\sum_{x\in\mathbb{Z}^4}\,
G(x)e^{-ip.x}$, where $p=(p^0,\vec p)$ are conjugate variables, has
a lattice K\"allen-Lehman type representation which allows us to
relate momentum space singularities on the imaginary $p^0$ axis to
points in the E-M spectrum. We want to show that there are isolated
dispersion curves, defining the eightfold way baryons and their
anti-baryons, up to near the energy threshold of $-5\ln\kappa$. To
this end, we consider the inverse $\tilde\Gamma(p)=\tilde
G(p)^{-1}$, and show that, for fixed $\vec p$ and $\kappa$,
$$\tilde\Gamma^{-1}(p)= {\{{\rm cof}\,[ \tilde\Gamma(p)]\}^t}/{{\rm
det}\tilde\Gamma(p)}\,,$$ provides a meromorphic extension of
$\tilde G(p)$ in $p^0$. Thus, the singularities of $\tilde G(p)$, or
spectral points, are contained in the zeroes of ${\rm
det\,}\tilde\Gamma(p)$ and the dispersion curves $w(\vec p)$ satisfy
the equation \bequ\lb{dispeq}{\rm det\,}\tilde \Gamma(p^0=iw(\vec
p), \vec p,\kappa)=0\,.\eequ That $\tilde\Gamma(p)^{-1}$ provides a
meromorphic extension of $\tilde G(p)$ follows from the faster
temporal falloff of $\Gamma(x=u-v)$, the convolution inverse of $G$,
as compared to $G$. The faster falloff of $\Gamma(x)$ gives us a
larger strip of analyticity in $\Im p^0$ which is $|\Im p^0|\leq
-(5-\epsilon)\ln\kappa$, $0<\epsilon\ll 1$, and the analyticity
implies the zeros of ${\rm det}\tilde\Gamma(p)$ are isolated, for
each $\vec p$ and $\kappa$, leading to the existence of particles in
the E-M spectrum. To obtain precisely the umber and behavior of the
dispersion curves, we need the short distance, low $\kappa$ order
behavior of $\Gamma$ which in turn follows from that of $G$.

To find the appropriate two-point function $G$, the associated
fields that create the low-lying excitation spectrum, as well as the
falloff of $G$ and $\Gamma$, we use a hyperplane decoupling method.
In this method, we consider the correlation
\bequ\lb{fk1}G_{LM}(u,v)\equiv =\left\langle
L(u)M(v)\right\rangle\,,\eequ for $u^0\not=v^0$, and with $L$ and
$M$ containing an odd number of $\hat \psi$ fields. For each set of
adjacent temporal hyperplanes (parametrized by $p$), which separate
the points,  we substitute the action hopping parameter $\kappa$ by
complex $\kappa_p$. Concerning the decay of $G_{LM}(u,v)$,
intuitively we pick up a decay factor of $\kappa_p$ for each
vanishing $\kappa_p$ derivative at $\kappa_p=0$. Taking the
$\kappa_p$ derivatives at $\kappa_p=0$, the $r^{\rm th}$ derivatives
$r=0,1,2,4$ are zero by imbalance of fermions and or by
interhyperplane gauge field integration. To analyze the third
derivative, we need the gauge integral (see Refs.
\cite{Creu2,longo}) $\int g_{a_1b_1}g_{a_2b_2}g_{a_3b_3}d\mu(g)=
\epsilon_{a_1a_2a_3}\epsilon_{b_1b_2b_3}/6$. With all fields at the
same point, we let $$\hat b_{\vec \alpha\vec
f}=\;\epsilon_{abc}\,\hat\psi_{a\alpha_1f_1}\hat\psi_{b\alpha_2f_2}
\hat\psi_{c\alpha_3f_3}\,,$$ and use the superscript $u$ ($\ell$) to
denote that only upper, $\alpha_i=1,2$ (lower, $\alpha_i=3,4$) spin
components occur. Also, the superscript $^{(r)}$ means the
coefficient of $\kappa_p^r$. For $v^0>u^0$, $u^0+1/2\leq p\leq
v^0-1/2$, we obtain \bequ\lb{bar}\barr{lll}\langle
L(u)M(v)\rangle^{(3)}\!\!&=&\!\!-\frac 1{6^2}\sum_{\vec \gamma,\vec
g, \vec w} [ \langle L(u)\bar b^\ell_{\vec\gamma\vec g}(p,\vec
w)\rangle^{(0)}
\langle b^\ell_{\vec\gamma\vec g}(p+1,\vec
w)M(v)\rangle^{(0)}
-\langle L(u) b^u_{\vec\gamma\vec g}(p,\vec
w)\rangle^{(0)}
\langle \bar b^u_{\vec\gamma\vec g}(p+1,\vec
w)M(v)\rangle^{(0)}]\,.\earr\eequ A similar expression holds for
$u^0>v^0$. Note that, with the Levi-Civita's coming from ${\cal
I}_3$, the local, gauge-invariant (colorless) fields $\bar b_{\vec
\alpha\vec f}$ and $b_{\vec \alpha\vec f}$ have naturally made their
appearance. Below, we will show that $\bar b^\ell_{\vec \alpha\vec
f}$ ($b^u_{\vec \alpha\vec f}$) are the basic excitations. Their
linear combinations create the eightfold way baryons and
anti-baryons which are related by charge conjugation and have
identical spectral properties.

From the vanishing of the zeroth, the first and the second
$\kappa_p$ derivatives, which extends to spatial separations as
well, we obtain the decay \bequ\lb{bound}|G_{LM}(u,v)|\leq {\rm
const}\;\kappa^{3|u-v|}\,,\eequ with $|u-v|\equiv |u^0-v^0|+|\vec
u-\vec v|$, $|\vec u-\vec v|=\sum_{i=1,2,3}\,|u^i-v^i|$, and $\tilde
G(p)$ is analytic in the strip $|\Im p^0|<-(3-\epsilon)\ln\kappa$.

We emphasize that there is no guesswork regarding the form of the
baryon fields in Eq. (\ref{bar}), and in the sequel we only consider
these fields. The appropriate choice for the two-point correlation
in our method is to choose $L$ and $M$ so that the correlation on
the lhs of Eq. (\ref{bar}) is the same as those on the rhs, i.e. to
have {\em closure}. For this, we take $L=b_{\vec \alpha\vec f}$ and
$M=\bar b_{\vec \beta\vec h}$ with lower indices in $\vec \alpha$,
such that only the first term in Eq. (\ref{bar}) survives (choosing
$L=\bar b^u_{\vec \alpha\vec f}\,$, $M=b^u_{\vec \beta\vec h}$ only
the second term survives and leads to antibaryons). Now, with a view
to obtaining a convolution inverse with a faster temporal decay, we
note that there are redundancies in $\hat b_{\vec \alpha\vec f}$,
which prevents us from defining the inverse of $G_{LM}$. The linear
dependencies are eliminated using what we call the {\em totally
symmetric property} (tsp) which is invariance of $\hat b$ under the
exchanges $\alpha_i\,f_i\leftrightarrow \alpha_j\,f_j$. By
considering only one element per equivalence class we have only $56$
independent fields, instead of the original $6^3=216$ (see Ref.
\cite{Ham}). The number $56$ arises as it is the dimension of the
totally symmetric subspace of the $3-$fold product of a
$6$-dimensional vector space. The dimension $6$ comes from $3$
(flavors) $\times\: 2$ (lower spins). With this restriction in
effect and without changing notation we can now introduce the
normalized fields $\hat B_{\vec \alpha\vec f}=\hat b_{\vec
\alpha\vec f}/[n_{\vec \alpha\vec f}]$ such that, for coincident
points, $\langle B_{\vec \alpha\vec f}\bar B_{\vec
\alpha^{\,\prime}\vec f^{\,\prime}}\rangle^{(0)}=-\delta_{\vec
\alpha\vec \alpha^{\,\prime}}\delta_{\vec f\vec f^{\,\prime}}$. To
see the last equality we use the basic formula $\langle b_{\vec
\alpha\vec f}\,\bar b_{\vec \alpha^{\,\prime}\vec
f^{\,\prime}}\rangle^{(0)}=-6\;{\rm perm}({\cal A})$, where ${\rm
perm}$ is the permanent (similar to ${\rm det}$, but with only $+$
signs), and ${\cal A}_{ij}=
\delta_{\alpha_i\alpha^\prime_j}\,\delta_{f_if^\prime_j}$, for
$i,j=1,2,3$.

Taking into account the other time ordering, we define the
two-baryon function for all $u$ and $v$ by ($\chi$ is the
characteristic function here)
\bequ\lb{gggg}G_{\ell_1\ell_2}(u,v)=\langle B_{\ell_1}(u)\bar
B_{\ell_2}(v)\rangle\,\chi_{u^0\leq v^0}-\langle \bar B_{\ell_1}(u)
B_{\ell_2}(v)\rangle^*\,\chi_{u^0> v^0}\,,\eequ where now the
$\ell$'s are collective indices for $\vec\alpha\vec f$'s, and we
suppress the lower spin superscripts. For $u^0\not= v^0$,
$$G^{(3)}_{\ell_1\ell_2}(u,v)=-\sum_{\ell_3,\vec w}\,
G^{(0)}_{\ell_1\ell_3}(u,(p,\vec w))\,
G^{(0)}_{\ell_3\ell_2}((p+1,\vec
w),v)\chi_{u^0<v^0}-\sum_{\ell_3,\vec w}\,
G^{(0)}_{\ell_1\ell_3}(u,(p+1,\vec w))\,
G^{(0)}_{\ell_3\ell_2}((p,\vec w),v)\chi_{u^0>v^0}\,,$$ which we
write symbolically as \bequ\lb{circ}G^{(3)}(u,v)=-[G^{(0)}\circ
G^{(0)}](u,v)\,,\eequ with true convolution in space. This important
relation, which we call the {\em product structure}, is instrumental
in showing the faster temporal $\Gamma$ decay, as it feeds into the
formula for the $3$rd $\kappa_p$ derivative of $\Gamma$ at
$\kappa=0$. Indeed, letting $G=G_d+G_n$, where
$G_{d,\ell_1\ell_2}(u,v)=G_{\ell_1\ell_2}(u,u)\delta_{\ell_1\ell_2}\delta
_{uv}$ is the diagonal part of $G$, we define $\Gamma$ by the
Neumann $\Gamma\equiv (G_d+G_n)^{-1}=\sum_{k=0}^\infty\,
G_d^{-1}(-G_nG_d^{-1})^k$, which converges by the global bound on
$G$ of Eq. (\ref{bound}). Using $\Gamma G=1=G\Gamma$, the Leibniz
formula $\partial^r \Gamma'= \sum_{s=0}^{r-1}\, {\small \left(
\barr{c}
r\\s\earr\right)}\,\Gamma'\,\partial^{r-s}G\,\partial^s\Gamma'$
(with $\Gamma'\equiv-\Gamma$), the product structure of Eq.
(\ref{circ}) and the $\kappa_p$ derivatives of $G$, we obtain
$\Gamma^{(r=0,1,2)}(u,v)=0$, for $|u^0-v^0|\geq 1$,
$\Gamma^{(3)}(u,v)=-[\Gamma^{(0)}G^{(3)}\Gamma^{(0)}](u,v)=0$, if
$|u^0-v^0|>1$ and, by imbalance of fermions, $\Gamma^{(4)}(u,v)=0$,
$|u^0-v^0|>1$. From this, using joint analyticity and Cauchy
estimates, follows the faster decay bound
\bequ\lb{gbd}|\Gamma_{\ell_1\ell_2}(u,v)|\leq {\rm
const}\;|\kappa|^{3}\,|\kappa|^{5(|u^0-v^0|-1)+3|\vec u-\vec
v|}\quad ,\quad |u^0-v^0|\geq 1\,,\eequ and the rhs is replaced by
${\rm const}\;\kappa^{3|\vec u-\vec v|}$, if $u^0=v^0$. Hence,
$\tilde\Gamma(p)$ is analytic in the larger strip $|\Im
p^0|<-(5-\epsilon)\ln\kappa$.

To relate points in the E-M spectrum to singularities of $\tilde
G_{\ell_1\ell_2}(p)$, we first use the F-K formula to obtain a
spectral representation, with $\bar B_\ell\equiv\bar B_\ell(1/2,\vec
0)$ and $x=v-u$, $$G_{\ell_1\ell_2}(x)=-(\bar B_{\ell_1}, \check
T^{|x^0|}\check T^{\vec x}\bar B_{\ell_2})_{\cal H}=-\int_{-1}^1\,
\int_{{\mathbb T}^3}\,\!(\la^0)^{|x^0|-1}
 e^{-i\vec \la.\vec x} d_{\la}(\bar B_{\ell_1},{\cal
E}(\la^0,\vec \la)\bar B_{\ell_2})_{{\cal H}}\,,$$ for $x\in\Z^{4}$,
$x^0\not= 0$, and is an even function of $\vec x$ by parity
symmetry. For the fourier transform, after separating out the
$x^0=0$ contribution, we get \bequ\lb{specrep} \barr{lcl}\tilde
G_{\ell_1\ell_2}(p)&\!\!=\!\!&\tilde G_{\ell_1\ell_2}(\vec p)\!
-\!(2\pi)^3
\!\dis\int_{-1}^{1} f(p^0,\la^0)d_{\la^0}\alpha_{\vec
p,\ell_1\ell_2}(\la^0)\,,\earr\eequ with $f(x,y)\equiv
(e^{ix}-y)^{-1}+(e^{-ix}-y)^{-1}$, where $d_{\la^0}\alpha_{\vec
p,\ell_1\ell_2}(\la^0)\!=\!\int_{\mathbb{T}^3}\delta(\vec p-\vec
\la)$ $d_{\la^0}d_{\vec\la}(\bar B_{\ell_1},{\cal E}(\la^0,\vec
\la)\bar B_{\ell_2})_{\cal H}$, and we have set $\tilde G (\vec
p)=\sum_{\vec x}\,e^{-i\vec p.\vec x}G(x^0=0,\vec x)$.

Singularities on the $\Im\,p^0$ axis are spectral points and are
contained in the zeroes of ${\rm det} \,\tilde \Gamma(\vec p)$. We
first restrict our attention to the determination of the masses
(i.e. $\vec p=\vec 0$), which is simplified passing to a basis where
$\tilde \Gamma(p^0,\vec p=\vec 0)$ is diagonal. The diagonalization
is achieved by fully exploiting the ${\rm SU}(3)_f$ symmetry, and
passing to the eightfold way baryon particle basis. The particle
basis is related to the individual spin and isospin basis we have
dealt with up to now by a linear real orthonormal transformation.
This transformation preserves the product structure, and thus the
larger strip analyticity region of $\tilde \Gamma_{r_1r_2}$, where
now the $r_i$'s are collective indices of the particle basis.

The use of the flavor symmetry reduces $\tilde \Gamma(\vec p)$ to a
block form with $8$ identical $2\times 2$ blocks associated with the
spin $1/2$ octet, and $10$ identical $4\times 4$ blocks associated
with the spin $3/2$ decuplet. The octet basis is given by
$$\barr{lll}p_\pm=\frac{\large \epsilon_{abc}}{3\sqrt{2}} (\bar\psi_{a+
u}\bar\psi_{b- d}-\bar\psi_{a+ d} \bar\psi_{b- u})\bar\psi_{c\pm
u}&,& n_\pm=\frac{\large \epsilon_{abc}}{3\sqrt{2}} (\bar\psi_{a+
u}\bar\psi_{b- d}-\bar\psi_{a+ d} \bar\psi_{b- u})\bar\psi_{c\pm
d},\\\Xi^0_\pm=\frac{\large \epsilon_{abc}}{3\sqrt{2}} (\bar\psi_{a+
u}\bar\psi_{b- s}-\bar\psi_{a+ s} \bar\psi_{b- u})\bar\psi_{c\pm
s}&,& \Xi^-_\pm=\frac{\large \epsilon_{abc}}{3\sqrt{2}}
(\bar\psi_{a+ d}\bar\psi_{b- s}-\bar\psi_{a+ s} \bar\psi_{b-
d})\bar\psi_{c\pm s},\\\Sigma^+_\pm=\frac{\large
\epsilon_{abc}}{3\sqrt{2}} (\bar\psi_{a+ u}\bar\psi_{b-
s}-\bar\psi_{a+ s} \bar\psi_{b- u})\bar\psi_{c\pm u}&,&
\Sigma^0_\pm=\frac{\large \epsilon_{abc}}{6} (2 \bar\psi_{a\pm
u}\bar\psi_{b\pm d}\bar\psi_{c\mp s}-\bar\psi_{a-u}\bar\psi_{b+
d}\bar\psi_{c\pm s}-\bar\psi_{a+u}\bar\psi_{b-d}\bar\psi_{c\pm
s}),\\
\Sigma^-_\pm=\frac{\large \epsilon_{abc}}{3\sqrt{2}} (\bar\psi_{a+
d}\bar\psi_{b- s}-\bar\psi_{a+ s} \bar\psi_{b- d})\bar\psi_{c\pm
d}&,& \Lambda_\pm=\frac{\large \epsilon_{abc}}{2\sqrt{3}}
(\bar\psi_{a+ u}\bar\psi_{b- d}-\bar\psi_{a+ d} \bar\psi_{b-
u})\bar\psi_{c\pm s},\earr$$ and the decuplet basis is given by
$$\barr{lll}\Delta^+_{\small \frac {\pm 1} 2}=\frac{\large \epsilon_{abc}}{6}
(\bar\psi_{a\pm u}\bar\psi_{b\pm u}\bar\psi_{c \mp
d}+2\bar\psi_{a\pm u}\bar\psi_{b\mp u}\bar\psi_{c \pm d})&,&
\Delta^+_{\small \frac{\pm3}2}=\frac{\large
\epsilon_{abc}}{2\sqrt{3}}\, \bar\psi_{a\pm u}\bar\psi_{b\pm
u}\bar\psi_{c\pm d},\\\Delta^0_{\small \frac {\pm 1} 2}=\frac{\large
\epsilon_{abc}}{6} (2\bar\psi_{a\pm u}\bar\psi_{b\pm
d}\bar\psi_{c\mp d}+\bar\psi_{a\mp u}\bar\psi_{b\pm d}\bar\psi_{c\pm
d})&,& \Delta^0_{\small \frac{\pm3}2}=\frac{\large
\epsilon_{abc}}{2\sqrt{3}}\, \bar\psi_{a\pm u}\bar\psi_{b\pm
d}\bar\psi_{c\pm d},\\ \Delta^-_{\small \frac {\pm1}2}=\frac{\large
\epsilon_{abc}}{2\sqrt{3}}\, \bar\psi_{a\pm d}\bar\psi_{b\pm
d}\bar\psi_{c\mp d}&,& \Delta^-_{\small \frac{ \pm3}2}=\frac{\large
\epsilon_{abc}}{6}\, \bar\psi_{a\pm d}\bar\psi_{b\pm
d}\bar\psi_{c\pm d},\\ \Delta^{++}_{\small \frac
{\pm1}2}=\frac{\large \epsilon_{abc}}{2\sqrt{3}}\, \bar\psi_{a\pm
u}\bar\psi_{b\pm u}\bar\psi_{c\mp u}&,& \Delta^{++}_{\small \frac{
\pm3}2}=\frac{\large \epsilon_{abc}}{6}\, \bar\psi_{a\pm
u}\bar\psi_{b\pm u}\bar\psi_{c\pm u},\\ \Sigma^{*+}_{\small \frac
{\pm3}2}=\!\frac{\large \epsilon_{abc}}{2\sqrt{3}} \bar\psi_{a\pm
u}\bar\psi_{b\pm u}\bar\psi_{c\pm s}&,& \Sigma^{*+}_{\small
\frac{\pm1}2}=\frac{\large \epsilon_{abc}}{6}\, (\bar\psi_{a\pm
u}\bar\psi_{b\pm u}\bar\psi_{c\mp s}+2\bar\psi_{a\pm
u}\bar\psi_{b\mp u}\bar\psi_{c\pm s}),\\ \Sigma^{*0}_{\small \frac{
\pm3}2}=\frac{\large \epsilon_{abc}}{6}\, \bar\psi_{a\pm
u}\bar\psi_{b\pm d}\bar\psi_{c\pm s}&,& \Sigma^{*0}_{\small \frac{
\pm1}2}= \!\frac{\large \epsilon_{abc}}{3\sqrt{2}} (\bar\psi_{a\pm
u}\bar\psi_{b\pm d}\bar\psi_{c\mp s} +\bar\psi_{a\pm
u}\bar\psi_{b\mp d}\bar\psi_{c\pm s} +\bar\psi_{a\mp
u}\bar\psi_{b\pm d}\bar\psi_{c\pm s}),\\ \Sigma^{*-}_{\small \frac
{\pm3}2}=\!\frac{\large \epsilon_{abc}}{2\sqrt{3}} \bar\psi_{a\pm
d}\bar\psi_{b\pm d}\bar\psi_{c\pm s}&,& \Sigma^{*-}_{\small
\frac{\pm1}2}=\frac{\large \epsilon_{abc}}{6} (\bar\psi_{a\pm
d}\bar\psi_{b\pm d}\bar\psi_{c\mp s}+2\bar\psi_{a\pm d}
\bar\psi_{b\mp d}\bar\psi_{c\pm s}),\\ \Xi^{*0}_{\small \frac
{\pm3}2}=\!\frac{\large \epsilon_{abc}}{2\sqrt{3}} \bar\psi_{a\pm
u}\bar\psi_{b\pm s}\bar\psi_{c\pm s}&,& \Xi^{*0}_{\small
\frac{\pm1}2}=\frac{\large \epsilon_{abc}}{6} (\bar\psi_{a\mp
u}\bar\psi_{b\pm s}+2\bar\psi_{a\pm u}\bar\psi_{b\mp
s})\bar\psi_{c\pm s},\\ \Xi^{*-}_{\small \frac
{\pm3}2}=\!\frac{\large \epsilon_{abc}}{2\sqrt{3}} \bar\psi_{a\pm
d}\bar\psi_{b\pm s}\bar\psi_{c \pm s}&,& \Xi^{*-}_{\small
\frac{\pm1}2}=\frac{\large \epsilon_{abc}}{6}\, (\bar\psi_{a\mp
d}\bar\psi_{b\pm s}+2\bar\psi_{a\pm d}\bar\psi_{b\mp
s})\bar\psi_{c\pm s},\\ \Omega^{-}_{\small \frac
{\pm3}2}=\!\frac{\large \epsilon_{abc}}{6} \bar\psi_{a\pm
s}\bar\psi_{b\pm s}\bar\psi_{c \pm s}&,& \Omega^{-}_{\small
 \frac{\pm1}2}=\!\frac{\large \epsilon_{abc}}{2\sqrt{3}}\bar\psi_{a\pm s}
 \bar\psi_{b\pm s}\bar\psi_{c\mp s}\,.\earr$$ We use the same
 particle symbol on the left, as for the G안ll-Mann-Ne안man eightfold way in the
 continuum, and the barred fields that creates it at the right. The
 particle basis collective indices specify the quantum numbers of total isospin $I$,
 third component $I_3$, the total hypercharge $Y$ and the value of
 the quadratic Casimir $C_2$ ($C_2=3$ for the octet and $6$
 for the decuplet.), besides the labels of total spin $J$
 and the $z$-component $J_z$. $J_z$ is the particle symbol
subscript. These labels are the same as the conventional ones
 for the continuum found in e.g. Refs. \cite{GellNee,Griffi}. For
 fixed $J_z$, the octet (decuplet) vectors form a basis for the $8$
 ($10$) dimensional irreducible representation of ${\rm SU}(3)_f$.
 Here we naturally obtain one $8$ and one $10$ dimensional
 representation of ${\rm SU}(3)_f$, which is not an {\em ad hoc} choice as in the
 $3\times 3 \times 3=10\oplus 8\oplus 8\oplus 1$ decomposition
 appearing in the non-dynamical group theoretical construction.

 The $j$-th component of total isospin, for $j=1,2,3$,
is defined through the linear operator acting on  a function $F$ of
the Grassmann algebra by $A_j F\equiv\lim_{\theta\searrow 0}
 \{[F(\{U_j\bar \psi\},\{\psi U_j^\dag\})-F(\{\bar
 \psi\},\{\psi\})]/(i\theta)\}$, where $U_j\equiv
 U_j(\theta)=\exp(i\lambda_j\theta/2)$, $j=1, ..., 8$ is an element of
 ${\rm SU}(3)_f$ and the $\lambda_j$ are the usual G안ll-Mann matrices
\cite{GellNee,Griffi}. $Y$ is defined as $2A_8/\sqrt{3}$ and
$C_2=\sum_{j=1,...,8}A_j^2$. Defining the linear operator
$W(U)F=F(\{U\bar \psi\},\{\psi U^\dag\})$ then $W(U)$ lifts to a
unitary operator $\check W(U)$ on ${\cal H}$ by using the F-K
formula and the ${\rm SU}(3)_f$ symmetry. The generators $\check A_j
\equiv\lim_{\theta\searrow 0}
 \{\check W(U_j)-1]/(i\theta)\}$ of the eight one-parameter
subgroups are self-adjoint operators in ${\cal H}$. $\check I_3$,
$\check I^2=I_1^2+I_2^2+I_3^2$, $\check Y$, $\check C_2$ are
mutually commuting and their eigenvalues are quantum numbers which
are used to label the states. $J_x$, $J_y$, and $J_z$ are also
defined similarly on the Grassmann algebra only, with $U=U_2\oplus
U_2$, $U_2=\exp(i\sigma^j\theta/2)\in {\rm SU}(2)$. Of course,
formally taking the continuum limit, the generator $J_z$ is the
usual $z$-component generator for rotations of spin. The eigenvalues
of $J_z$ and $\vec J^{\,2}$ are also used to label the states.

We now use the auxiliary function method to determine: the octet
mass (they are all the same), the decuplet masses (they are all the
same up to and including ${\cal O}(\kappa^6)$), and the ${\cal
O}(\kappa^6)$ octet-decuplet mass splitting. By the use of the
symmetry of $\pi/2$ rotations about $e^3$, the matrix $\tilde
\Gamma(\vec p)$ is seen to be diagonal at $\vec p=\vec 0$; and by
$e^1$ reflections the elements only depend on $|J_z|$. The
determinant factorizes, and we consider one of the $56$ typical
factors (for which we omit all indices). As the mass runs out to
infinity as $\kappa\searrow 0$, the usual implicit function theorem
does not apply to solve Eq. (\ref{dispeq}) at $\vec p=\vec 0$. We
make a nonlinear transformation from $p^0$ to an auxiliary variable
\bequ\lb{wp0}w=-1-c_3(\vec p)\kappa^3+\kappa^3e^{-ip^0}\,,\eequ with
$c_3(\vec p)=-\sum_{j=1,2,3}\,\cos p^j/4$, and introduce an
auxiliary function $H(w,\kappa)$ such that $\tilde\Gamma(p^0,\vec
p)=H(w=-1-c_3(\vec p)\kappa^3+\kappa^3e^{-ip^0},\kappa)$. By doing
this, we bring the non-singular part of the mass $M+3\ln\kappa$ from
infinity to close to $w=0$, as $\kappa\searrow 0$. Using time
reversal and parity symmetry, we have $\Gamma(x^0,\vec
x)=\Gamma(-x^0,\vec x)$, and $H(w,\kappa)$ is defined by
\bequ\lb{HH}H(w,\kappa)=\sum_{\vec x}
\Gamma(x^0=0,\vec x)\,e^{-i\vec p.\vec x}+
\sum_{\vec x,n=1,2\ldots}\,\Gamma(n,\vec x)\left[ \left(
\frac{1+w+c_3(\vec p)\kappa^3}{\kappa^3}\right)^n +\left(
\frac{\kappa^3}{1+w+c_3(\vec
p)\kappa^3}\right)^{n\,}\right]\,e^{-i\vec p.\vec x}\,. \eequ The
global bound on $\Gamma$ of Eq. (\ref{gbd}) guarantees that
$H(w,\kappa)$ is jointly analytic in $\kappa$ and $w$. However, we
want to control the mass to order $\kappa^6$ and for this we need
the low $\kappa$ order short distance behavior of $\Gamma(x)$, which
in turn follows from that of $G(x)$. More precisely, we need
$\Gamma(x^0=n,\vec x)/\kappa^{3n}$) up to and including order
$\kappa^6$. The normalization condition $G^{(0)}(x=0)=-1$ implies
$\Gamma^{(0)}(x=0)=-1$ and, by a simple argument, the product
formula gives $G(x=e^0)=-\kappa^3+{\cal O}(\kappa^4)$, which implies
$\Gamma(x=e^0)=\kappa^3+{\cal O}(\kappa^4)$. Other contributions are
found by explicit calculation of coefficients of the hopping
parameter expansion of $G(x)$. Namely, there are contributions
arising from non-intersecting paths connecting the point $0$ to $x$
and paths that emit and absorb a meson. After a lengthy calculation,
we find $H(w,\kappa)=w +\frac{\kappa^6}{1+w}+a_6\kappa^6+b
\kappa^6+\kappa^6\sum_{n=1,...,4}\,c_{3n+6}^\prime (1+w)^n+
h(w,\kappa) \kappa^7$, with $b$ and the $c^\prime$압 taking the same
value for the octet and the decuplet, and $h(w,\kappa)$ jointly
analytic in $w$ and $\kappa$. Note that, in the term $a_6\kappa^6$,
we have separated the contributions coming from all points of the
form $x=\epsilon e^i+\epsilon^\prime e^j$, with $ij=12,13,23$,
$\epsilon,\epsilon^\prime=\pm1$, which we call {\em spatial angles}.
$a_6$ takes the value $a_o=3/8$ ($a_d=-3/8$) for the octet
(decuplet). As $H(0,0)=0$ and $[\partial H/\partial w](0,0)=1$, the
analytic implicit function theorem implies that $H(w,\kappa)=0$ has
the analytic solution $w(\kappa)$ given by, with $b^\prime=b+1 +
\sum_{n=1,...,4}\,c_{3n+6}^\prime$,
$$w(\kappa)=-a_6-b^\prime\kappa^6+{\cal O}(\kappa^7)\,.$$

Returning to Eq. (\ref{wp0}) and setting $p^0=iM$, we get
$$M=-\ln \kappa^3+\ln (1+c_3(\vec 0)
\kappa^3+w(\kappa))
=-3\ln\kappa+c_3(\vec 0)\kappa^3-[a_6+b^\prime+c_3(\vec
0)^2/2]\kappa^6+{\cal O}(\kappa^7)\,,$$ and the octet-decuplet mass
difference is $M_d-M_o=(a_o-a_d)\kappa^6+{\cal O}(\kappa^7)=
3\kappa^6/4+{\cal O}(\kappa^7)$.

For the $\vec p\not=\vec 0$ dispersion curves, the $2\times 2$ and
the $4\times 4$ blocks of $\tilde\Gamma(p)$ still have a complicated
structure even after the use of the usual well known symmetries.
However, we have found a new local symmetry of spin flip ${\cal
F}_s$ which is a composition of the nonlocal, antilinear time
reversal, the local, linear charge conjugation and a nonlocal,
linear time coordinate reflection \cite{treflec}. This symmetry
shows that the $2\times 2$ octet blocks are diagonal and a multiple
of the identity. For the octet, the identical dispersion curves
$w(\vec p)$ can be obtained by using the auxiliary function method
as done before for the masses. For the $4\times 4$ decuplet blocks,
the spin flip symmetry simplifies the matrix $\tilde\Gamma(\vec p)$
but it is {\em not} diagonal. We have not been able to apply the
auxiliary function method. However, we can use a Rouch\'e theorem
argument (principle of the argument) \cite{Si} on ${\rm det}
\tilde\Gamma_{4\times 4}(\vec p)$ to show that for each fixed $\vec
p$ there are exactly four pairwise identical solutions.

The dispersion relations (curves for the octet) admit the
representation $w(\kappa,\vec
p)=-3\ln\kappa-3\kappa^3/4+\kappa^3\sum_{j=1,2,3}\,(1-\cos
p^j)/4+r(\kappa,\vec p)$, where $r(\kappa,\vec p)$ is of ${\cal
O}(\kappa^6)$. For the octet, $r(\kappa,\vec p)$ is analytic in
$\kappa$ and in each $p^j$, for small $|\Im p^j|$.

Up to now we have shown the existence of baryons in the subspace
${\cal H}_b\subset {\cal H}_o$, the subspace generated by vectors of
the form $\hat b_{\vec\alpha\vec f}$. The E-M spectrum up to near
the meson-baryon threshold of $-5\ln\kappa$ consists of dispersion
relations associated with the $56$ eightfold way baryons and the
$56$ anti-baryons. In principle, there may be other states in ${\cal
H}_o$ with spectrum in the same energy interval. We show this is
{\em not} the case using a correlation subtraction method
\cite{CMP}.

We make some concluding remarks. We have considered a model with the
same quark mass and flavor ${\rm SU}(3)_f$ symmetry, which has
baryons and antibaryons. Our method extends to treat rigorously the
case of ${\rm SU}(N)_f$ to uncover $(2N+2)!/[3!(2N-1)!]$ baryons and
also to treat the case with different quark masses and broken flavor
symmetry. The product structure still holds, the elementary
excitations are revealed and are the same as the $\bar b_{\vec
\alpha\vec f}$ obtained here. The difficulty in obtaining the
dispersion curves for the decuplet particles disappears for the
continuum if the decuplet fields transform under the Poincar\'e
group. In this case, the four identical dispersion curves are of
course the relativistic ones.

The determination of the one-hadron spectrum is a necessary step to
analyze the existence of bound states. Our method is powerful enough
to access the hadron-hadron spectrum and should help in clarifying
fundamental open questions as e.g. the existence of certain
meson-baryon bound-states and pentaquarks.

This work was supported by CNPq and FAPESP.


\begin{thebibliography}{8fdw}
\bibitem {treflec} This interesting and new symmetry will be the subject
of a future publication.
\end{thebibliography}
\end{document}